\documentclass[twocolumn]{aastex62}

\received{2019 August 27}
\revised{2019 October 16}
\accepted{2019 October 21}
\published{2019 December 12}
\submitjournal{ApJS}

\begin{document}

\title{Lithium-rich Giants in LAMOST Survey. I. The Catalog}

\author{Qi Gao}
\altaffiliation{}
\affil{Key Laboratory of Optical Astronomy, National Astronomical Observatories, Chinese Academy of Sciences, Beijing 100101, People’s Republic of China}
\affil{University of Chinese Academy of Sciences, Beijing 100049, People’s Republic of China}

\author{Jian-Rong Shi}
\altaffiliation{E-mail: sjr@bao.ac.cn}
\affil{Key Laboratory of Optical Astronomy, National Astronomical Observatories, Chinese Academy of Sciences, Beijing 100101, People’s Republic of China}
\affil{University of Chinese Academy of Sciences, Beijing 100049, People’s Republic of China}

\author{Hong-Liang Yan}
\altaffiliation{E-mail: hlyan@nao.cas.cn}
\affil{Key Laboratory of Optical Astronomy, National Astronomical Observatories, Chinese Academy of Sciences, Beijing 100101, People’s Republic of China}
\affil{University of Chinese Academy of Sciences, Beijing 100049, People’s Republic of China}

\author{Tai-Sheng Yan}
\affil{Key Laboratory of Optical Astronomy, National Astronomical Observatories, Chinese Academy of Sciences, Beijing 100101, People’s Republic of China}
\affil{University of Chinese Academy of Sciences, Beijing 100049, People’s Republic of China}

\author{Mao-Sheng Xiang}
\affil{Max-Planck Institute for Astronomy, Konigstuhl, D-69117, Heidelberg, Germany}

\author{Yu-Tao Zhou}
\affil{Department of Astronomy, School of Physics, Peking University, Beijing 100871, China}
\affil{Kavli Institute for Astronomy and Astrophysics, Peking University, Beijing 100871, China}

\author{Chun-Qian Li}
\affil{Key Laboratory of Optical Astronomy, National Astronomical Observatories, Chinese Academy of Sciences, Beijing 100101, People’s Republic of China}
\affil{University of Chinese Academy of Sciences, Beijing 100049, People’s Republic of China}

\author{Gang Zhao}
\affil{Key Laboratory of Optical Astronomy, National Astronomical Observatories, Chinese Academy of Sciences, Beijing 100101, People’s Republic of China}
\affil{University of Chinese Academy of Sciences, Beijing 100049, People’s Republic of China}

\begin{abstract}

Standard stellar evolution model predicts a severe depletion of lithium (Li) abundance during the first dredge up process (FDU). Yet a small fraction of giant stars are still found to preserve a considerable amount of Li in their atmospheres after the FDU. Those giants are usually identified as Li-rich by a widely used criterion, A(Li) $ > 1.5$\,{\it dex}. A large number of works dedicated to searching for investigating this minority of the giant family, and the amount of Li-rich giants, has been largely expanded on, especially in the era of big data. In this paper, we present a catalog of Li-rich giants found from the Large Sky Area Multi-Object Fiber Spectroscopic Telescope (LAMOST) survey with Li abundances derived from a template-matching method developed for LAMOST low-resolution spectra. The catalog contains $10,535$ Li-rich giants with Li abundances from $\sim 1.5$\ to $\sim 4.9$\,{\it dex}. We also confirm that the ratio of Li-rich phenomenon among giant stars is about $1\%$--or more specifically, $1.29\%$--from our statistically important sample. This is the largest Li-rich giant sample ever reported to date, which significantly exceeds amount of all the reported Li-rich giants combined. The catalog will help the community to better understand the Li-rich phenomenon in giant stars. 

\end{abstract}

\keywords{stars: abundances --- stars: late-type --- stars: evolution --- stars: chemically peculiar --- stars: statistics}

\section{INTRODUCTION}

Fragile elements, such as lithium (Li), will be easily destroyed in the deep layers of stellar atmospheres, where the temperatures are usually as high as (if not higher than) millions of Kelvins. During the first dredge-up (FDU) process, matters circulate from the surface of a star to the bottom of its convective shell, bringing a large amount of lithium down into the deep layers where they can hardly survive. Thus the severe depletion of Li in the atmosphere of a giant star is the natural consequence of stellar evolution \citep{Iben1967a, Iben1967b}. Assuming an initial abundance of A(Li)\footnote{A(Li) $= \log[{\rm N_{Li}}/{\rm N_{H}} + 12$, where ${\rm N_{Li}}$ and ${\rm N_{H}}$ is the number density of lithium and hydrogen, respectively. } $= 3.3$\,{\it dex} for a main sequence star of approximately solar metallicity and mass above $\sim 1.4$ $M_{\odot}$, diluted for $\sim 60$ times due to FDU, its Li abundance will be below $1.5$\,{\it dex} when it finishes FDU.

The predicted depletion has been confirmed by a large number observations of giants \citep[][for example]{Brown1989, Lind2009, Liu2014a, Kirby2016}. However, \citeauthor{Wallerstein1982} reported a K giant with A(Li) up to $3.2$\,{\it dex} in \citeyear{Wallerstein1982}. Since then, about 600 giants with A(Li) over 1.5\,{\it dex} were reported with object-ID/positions and Li abundances \citep[e.g.,][]{Brown1989, Reddy2005, Kumar2011, Ruchti2011, Kirby2012, Martell2013, Adamow2014, Casey2016, Li2018, Smiljanic2018, Deepak2019, Zhou2019, Singh2019a, Singh2019b}. Furthermore, a number of Li-rich giants with special features have been found \cite[e.g.,][]{Kumar2009, Adamow2012, Silva Aguirre2014, Yan2018}.
In addition, methods of searching for Li-rich giants from low-resolution spectra were reported in different works \citep[e.g.,][]{Martell2013, Kumar2018a, Casey2019}. All of these efforts largely expanded the Li-rich family and provided observational constraints helping to understand how Li is enhanced in the evolved stars \cite[e.g,][]{Alexander1967, Cameron1971, Sackmann1999, Siess1999, Denissenkov2004, Charbonnel2010}, and even how Li is evolved in each scale of our Galaxy \cite[e.g.,][]{Fu2018, Cescutti2019,  Carlos2019}.

Although a considerable amount of Li-rich giants have been reported, they are still rare objects compared to huge amount of normal ones. 
The ratio of Li-rich to normal giants is very low. 
\citet{Brown1989} found that only $\sim 1.5\%$ of giants are Li-rich in nearby stars, and similar ratios were reported by \citet{Kumar2011} and \citet{Liu2014a}, etc. Observations of the Galactic bulge revealed a slightly lower ratio of $0.5\%-0.7 \%$ \citep{Gonzalez2009, Lebzelter2012}, and an analogy result was found for the Galactic thick-disk objects by \citet{Monaco2011}. The ratios estimated from large survey programs are $\sim 0.9\%$ from Gaia-ESO survey \citep{Casey2016, Smiljanic2018}, $\sim 0.8\%$ from RAVE sample \citep{Ruchti2011}, $\sim 2.0\%$ from PTPS data \citep{Adamow2014} and $\sim 0.2\%-0.3\%$ from SDSS and GALAH data\citep{Martell2013, Deepak2019}. 
Li-rich giants have been \emph{sporadically} reported due to their rareness in the past $\sim$\, 40 years. Although hundreds of Li-rich giants with object-IDs/positions and abundances available to the astronomy community for further studies, their data were usually obtained from different work, introducing tricky biases due to the diverse methods, samples, and data qualities, etc. 
For better understanding the Li-rich phenomenon in the evolved stars, a catalog of Li-rich giants identified by systematic and coherent method from massive spectroscopic survey program is thus essential. 

Large Sky Area Multi-Object Fiber Spectroscopic Telescope (LAMOST) survey \citep{Cui2012,Zhao2012} has finished its six-years of phase-I survey in low-resolution mode (R $\simeq 1,800$), and has begun its phase-II survey in a combination of low- and medium-resolution mode (R $\simeq 7,500$). The low-resolution spectra observed by LAMOST to date number over 10 million.
It is almost certain that large amount of Li-rich giants are hidden in this vast database. The scope of this study is to systematically search for Li-rich giants from LAMOST DR7 low-resolution spectra data and to derive the Li abundances by template-matching method. We present a catalog of Li-rich giants obtained from LAMOST including LAMOST-ID, position, effective temperature, surface gravity, metallicity, Li abundance, etc.

The paper is assembled as follow: In Section \ref{sec:SAMPLE}, we briefly describe the giant sample selected from the LAMOST low-resolution spectra. The method and procedure of deriving the Li abundances and error estimation are described in details in Section \ref{sec:METHOD}, and in Section \ref{sec:RESULT}, we present the results of our Li-rich sample. Finally, a short discussion and summary are given in the \ref{sec:SUMMARY}th Section.

\section{STELLAR SAMPLE}\label{sec:SAMPLE}

\begin{figure}[!t]
\includegraphics[scale=0.37]{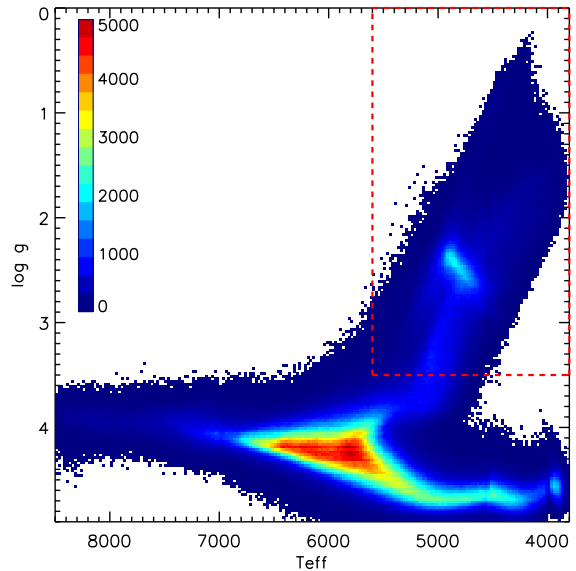}
\centering
\caption{The HR diagram of the stars observed by LAMOST low-resolution mode. The giant sample was identified by the following criteria: log \emph{g} {\textless} 3.5\,{\it dex} and {$\mathrm{\emph{T}_{eff}}$} {\textless} 5600\,K, as shown in the box with red dashed line.
\label{fig:1}}
\end{figure}

In this study, we used LAMOST low-resolution spectra obtained from October 2011 to June 2019, and the stellar atmospheric parameters ({$\mathrm{\emph{T}_{eff}}$}, log \emph{g}, [Fe/H]) and radial velocities (RV) determined by the LAMOST Stellar Parameter Pipeline \citep[LASP,][]{Luo2015}. The giants were selected based on the following criteria : log \emph{g} {\textless} 3.5 and {$\mathrm{\emph{T}_{eff}}$} {\textless} 5600\,K, which is revised from \citet{Liu2014b}. We got rid of the objects of 3.5 {\textless} log \emph{g} {\textless} 4.0 when 4600\,K {\textless} {$\mathrm{\emph{T}_{eff}}$} {\textless} 5600\,K which was also identified as giants by \citet{Liu2014b}, because they are contaminated by the newly formed stars with a little higher rate. The final sample includes 814,268 giants. Figure \ref{fig:1} shows the HR diagram of the all stellar objects observed by LAMOST low-resolution mode and the giant sample in the box with red dashed line.

\begin{figure*}[!t]
\includegraphics[scale=0.45]{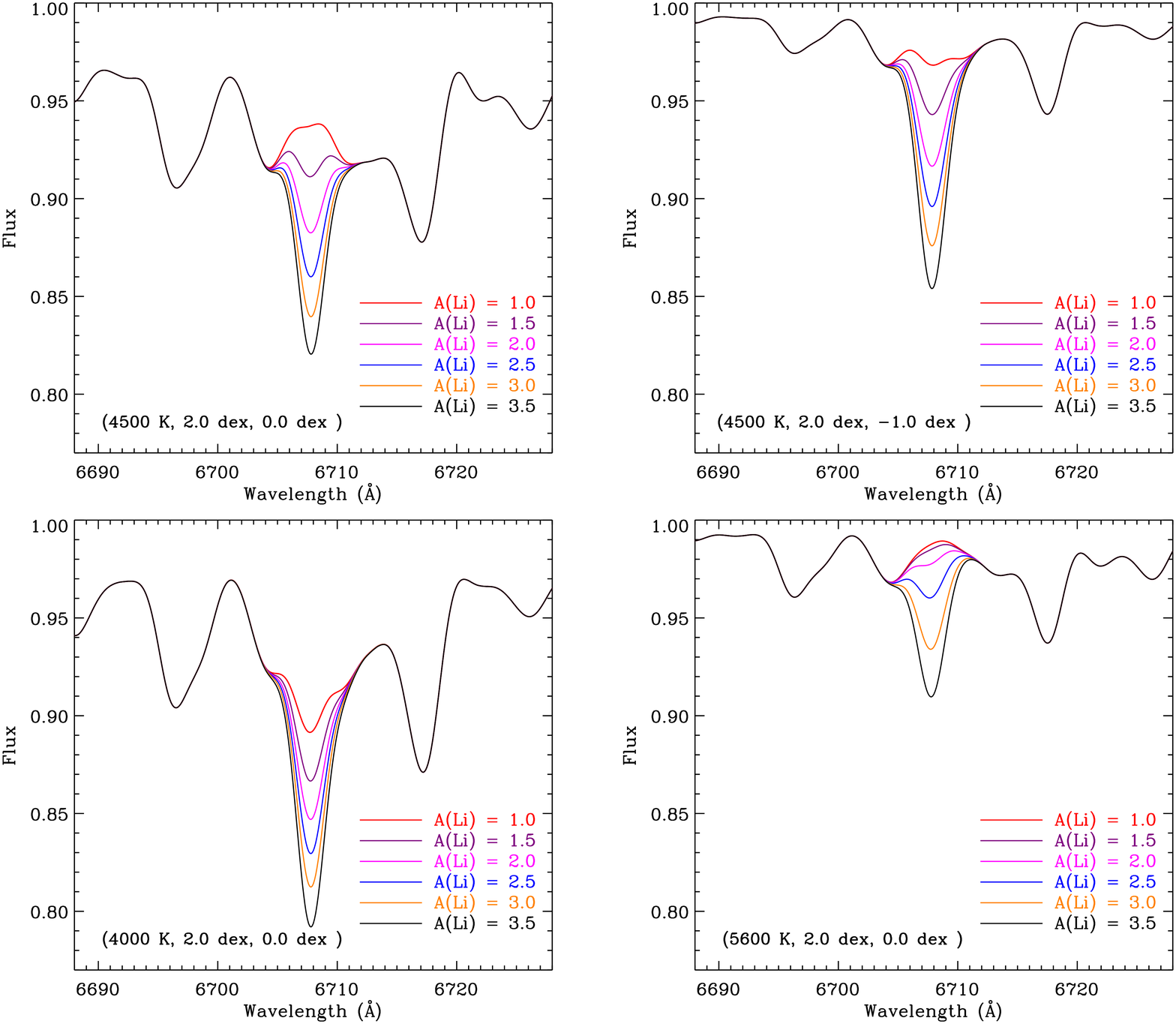}
\caption{The \ion{Li}{1} resonance lines varying with A(Li) from 1.0\,{\it dex} to 3.5\,{\it dex} in four sets of atmospheric parameters. The atmospheric parameters are marked in the sequence of {$\mathrm{\emph{T}_{eff}}$}, log \emph{g} and [Fe/H] and different A(Li) marked by different colors in each panel.\label{fig:2}}
\end{figure*}

\begin{figure*}[!t]
\includegraphics[scale=0.29]{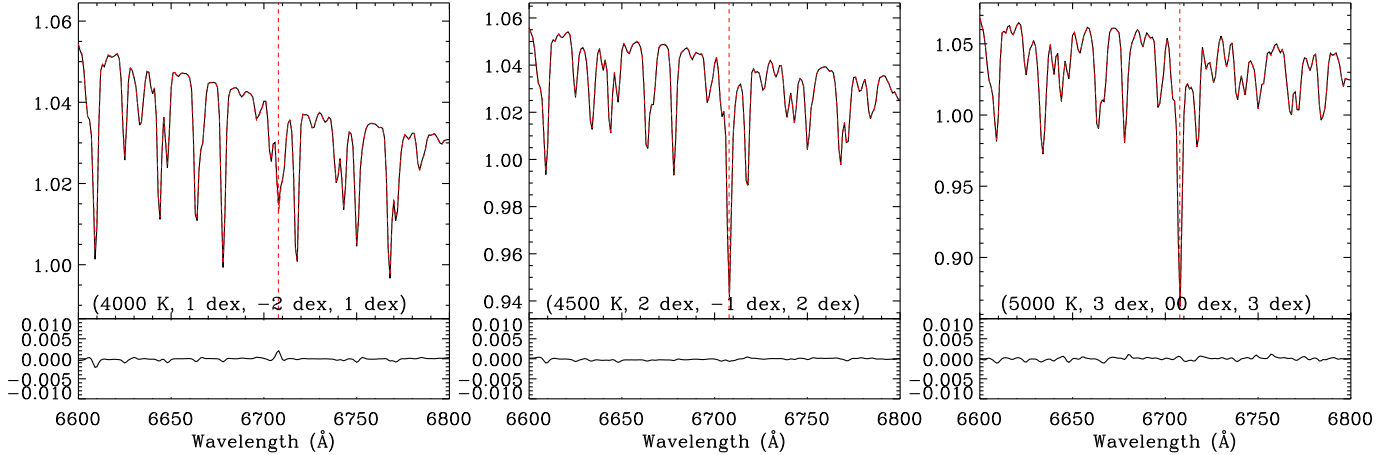}
\centering
\caption{Comparison of the original templates (red dashed line) and calculated ones interpolated from adjacent grids (black solid line) for three cases. Their atmospheric parameters ({$\mathrm{\emph{T}_{eff}}$}, log \emph{g}, [Fe/H] and [Li/Fe]) are presented, and the residual of the flux values are also plotted at the bottom of each panel.\label{fig:3}}
\end{figure*}

\begin{figure}[!t]
\includegraphics[scale=0.35]{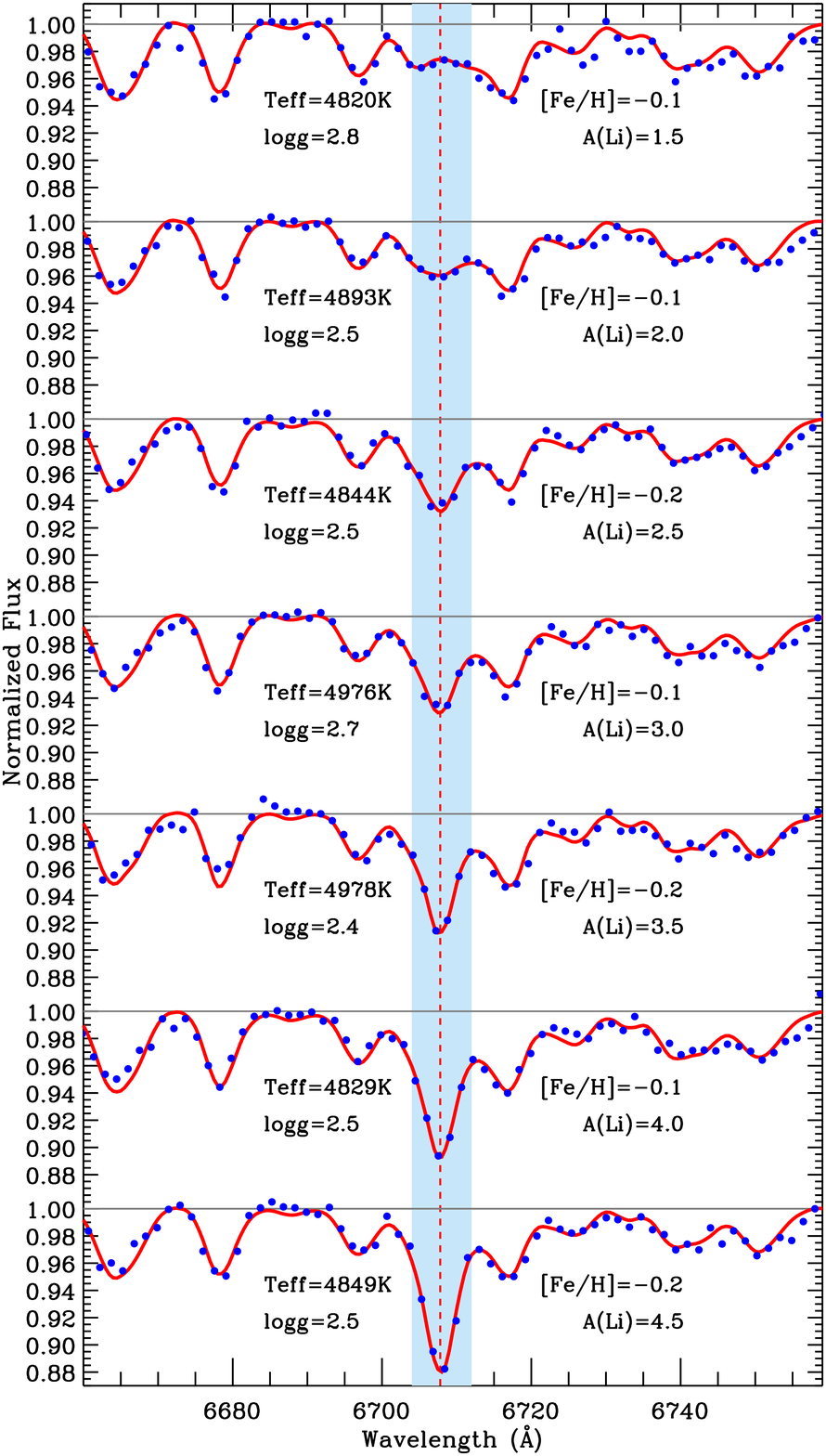}
\centering
\caption{Several fitting examples. The light blue region is the wavelength range to calculate $\chi^2$ value (6704\,-\,6712\,{\AA}), the blue dots represent the observed spectrum, and the red solid line is the best-matching template.\label{fig:4}}
\end{figure}

\begin{figure*}[!t]
\includegraphics[scale=0.35]{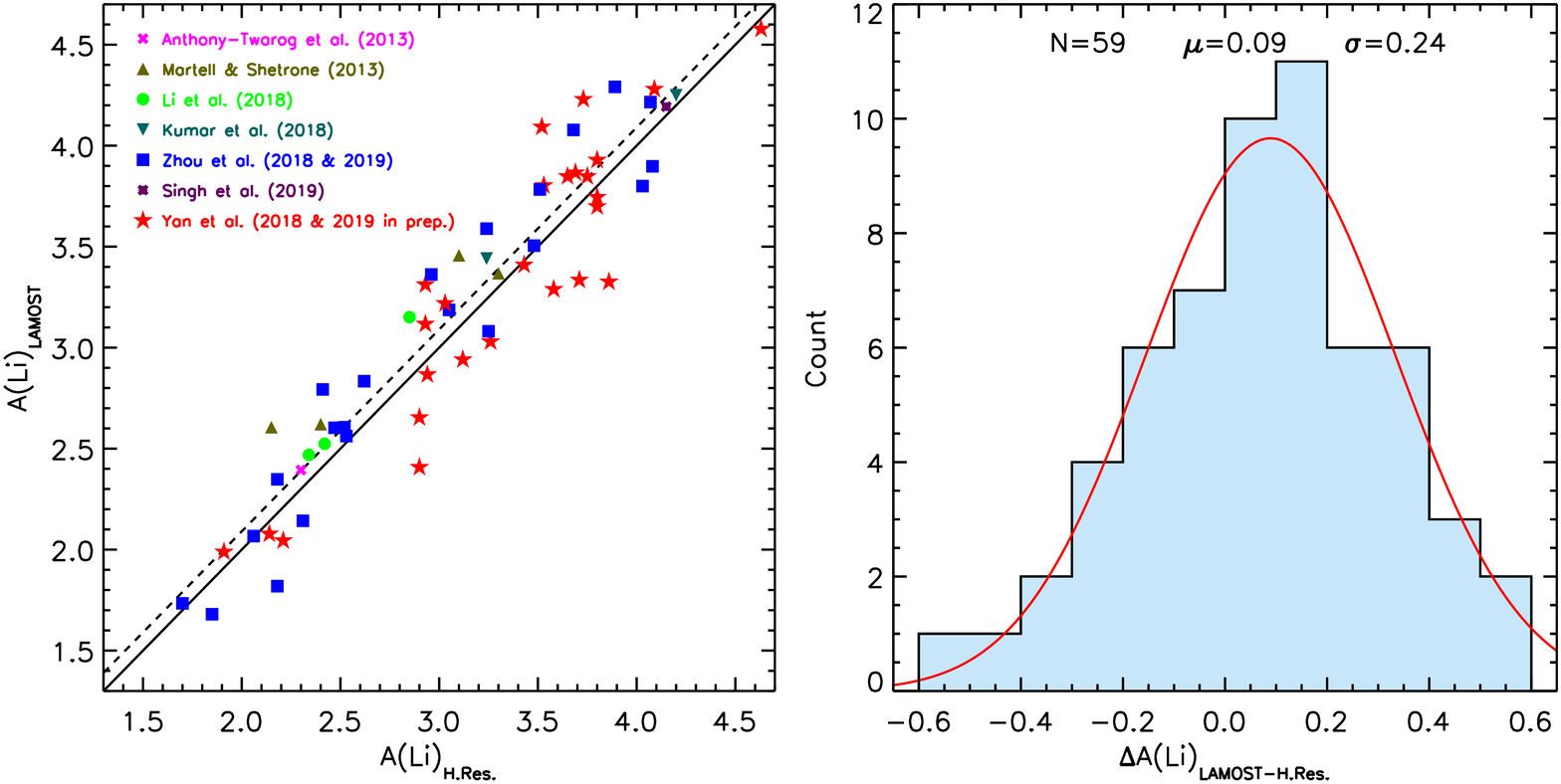}
\centering
\caption{The comparison of A(Li)$_{\rm LAMOST}$ derived by our method using the stellar atmospheric parameters provided in the literature and A(Li)$_{\rm H.Res.}$ provided in the literature. The left panel plots A(Li)$_{\rm LAMOST}$ against A(Li)$_{\rm H.Res.}$ with the one-to-one correspondence solid line and the corrected dashed line after an overall shift of 0.09\,{\it dex}, different markers and colors stand for different reporters. The right panel shows the distribution of differences between A(Li)$_{\rm LAMOST}$ and A(Li)$_{\rm H.Res.}$ and the result of Gaussian fit over-plotted with red solid line. \label{fig:5}}
\end{figure*}

\section{METHOD}\label{sec:METHOD} 

A template matching method has been adopted to determine the Li abundances in the term of [Li/Fe], then they are converted into the expression of A(Li) by the relationship of A(Li) = [Li/Fe] + [Fe/H] + A(Li){$_\odot$}. Our method of deriving the [Li/Fe] is similar to the method adopted by \citet{Li2016}, which is based on LSP3 \citep{Xiang2015} and was developed to determine the [$\alpha$/Fe] from LAMOST low-resolution spectra.

\subsection{The synthetic template spectra} 

The SPECTRUM synthesis code (V2.76, 2010) based on the Kurucz ODFNEW atmospheric models \citep{Castelli2003} with the standard abundance distribution of \citet{Grevesse1998} was used to calculate the template spectra. We applied the atomic line data of Li presented by \citet{shi2007}. In our calculations, a fixed micro-turbulence of 1.5 kms$^{-1}$ and a resolution of 2\,{\AA} have been adopted for all template spectra. The resolution of the LAMOST spectra is approximately 2.8 {\AA} on average and varies with each individual fiber \citep{Xiang2015}. Templates will be degraded in resolution according to each observed spectrum before matching.

We set the grids as follows: 3800\,K $\leq$ {$\mathrm{\emph{T}_{eff}}$} $\leq$ 5600\,K in steps of 100K, 0.0 $\leq$ log \emph{g} $\leq$ 4.0 in steps of 0.25\,{\it dex}, -2.6 \textless [Fe/H] $\leq$ 0.4 in steps of 0.2\,{\it dex} and -3.0 $\leq$ [Li/Fe] $\leq$ 6.9 in steps of 0.1\,{\it dex}. As the \ion{Li}{1} resonance line at 6708 {\AA} mixed with the nearby \ion{Ca}{1} line at 6717\,\AA\ for fast rotation stars, we took account of the influence of the $\alpha$-enhancement: the $\alpha$-element abundances enhanced by 0.4\,{\it dex} for stars of [Fe/H] $\textless$ -0.6\,{\it dex}. The \ion{Li}{1} resonance lines varying with A(Li) from 1.0\,{\it dex} to 3.5\,{\it dex} in four sets of atmospheric parameters are presented in Figure \ref{fig:2}.

\begin{figure}[!t]
\includegraphics[scale=0.37]{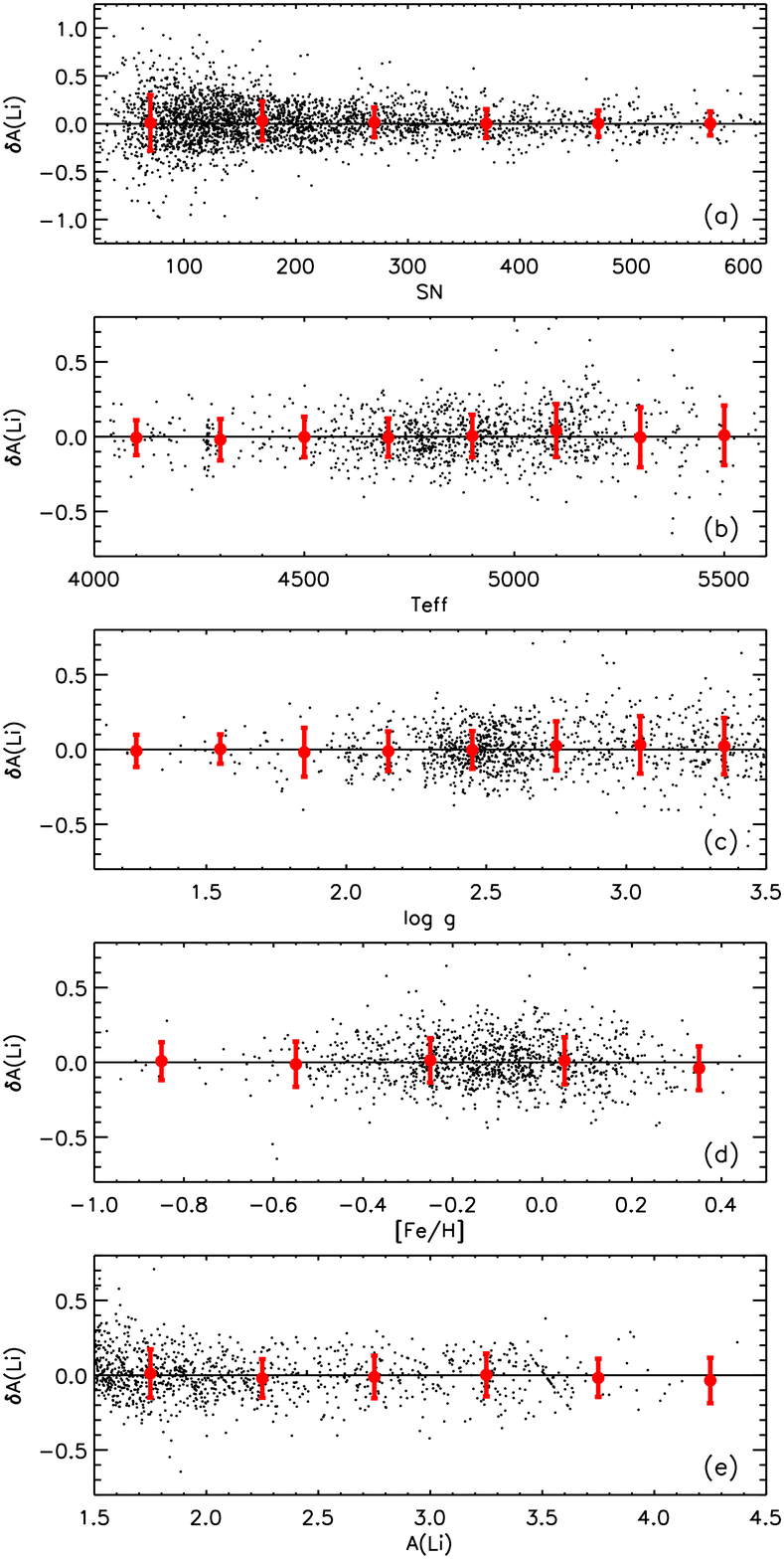}
\centering
\caption{The differences in A(Li) between multiple observations versus S/N, {$\mathrm{\emph{T}_{eff}}$}, log \emph{g}, [Fe/H] and A(Li). All the $2,746$ objects with repeated observations are included in panel (a), only $1,118$ objects with high quality data (S/N $\geq$ 200) were used in the rest panels.} The bin sizes are 100, 200\,K, 0.3\,{\it dex}, 0.3\,{\it dex} and 0.5\,{\it dex}, respectively. The red dots are the mean value and the error bars are the  standard deviation of the differences in every bin.\label{fig:6}
\end{figure}

\begin{figure*}[!t]
\includegraphics[scale=0.3]{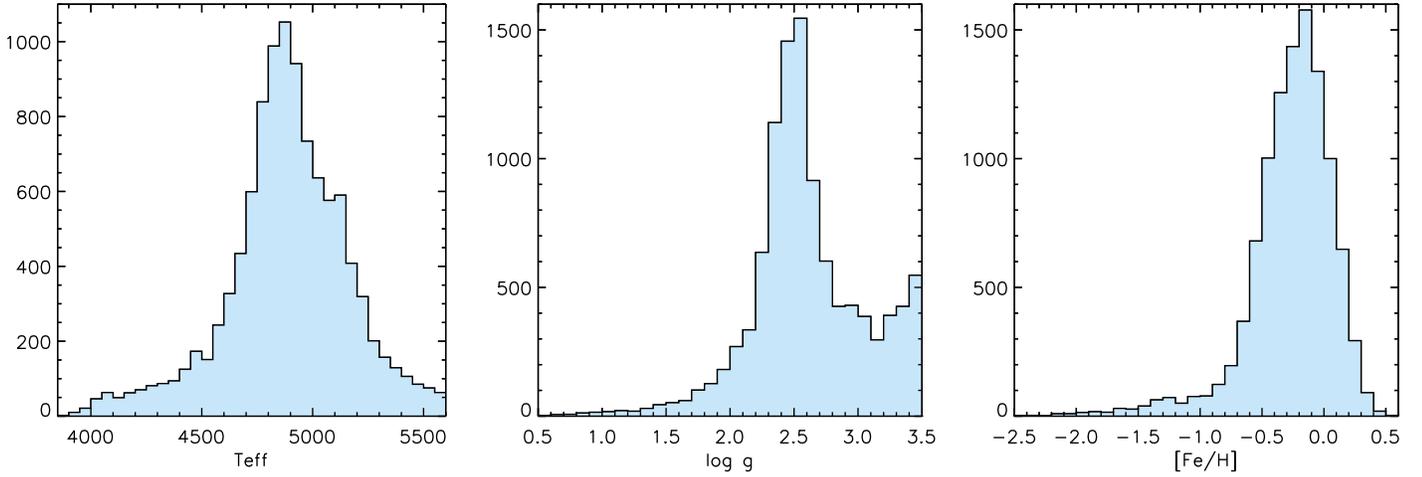}
\centering
\caption{Histograms of our Li-rich giant sample as functions of {$\mathrm{\emph{T}_{eff}}$}, log \emph{g} and [Fe/H]. The bin sizes are 50\,K, 0.1\,{\it dex} and 0.1\,{\it dex}, respectively. The distribution of Li-rich giants in temperature shows two peaks around 4800\,K and 5100\,K. For surface gravity, there are two clear peaks around log \emph{g}$\sim$2.5\,{\it dex} and 3.5\,{\it dex}.  And the distribution of our Li-rich sample in metallicity shows a clear peak around [Fe/H]$\sim$ -0.15\,{\it dex}. For metal poor stars there may be a second peak around $-1.25$\,{\it dex}.\label{fig:7}}
\end{figure*}

\begin{figure}[!t]
\includegraphics[scale=0.4]{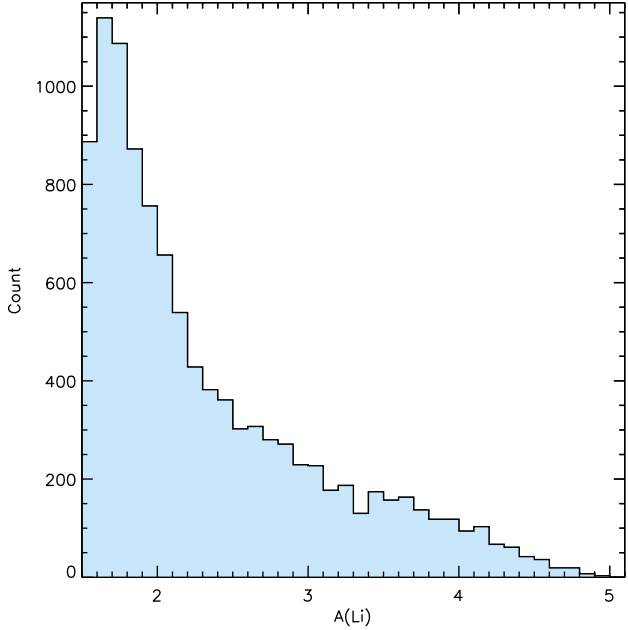}
\centering
\caption{Histogram of A(Li) of our Li-rich giant sample. The bin size is 0.1\,{\it dex}. The number of lithium giants for each bin declines with increasing A(Li) except the first bin. \label{fig:8}}
\end{figure}

\begin{figure}[!t]
\includegraphics[scale=0.4]{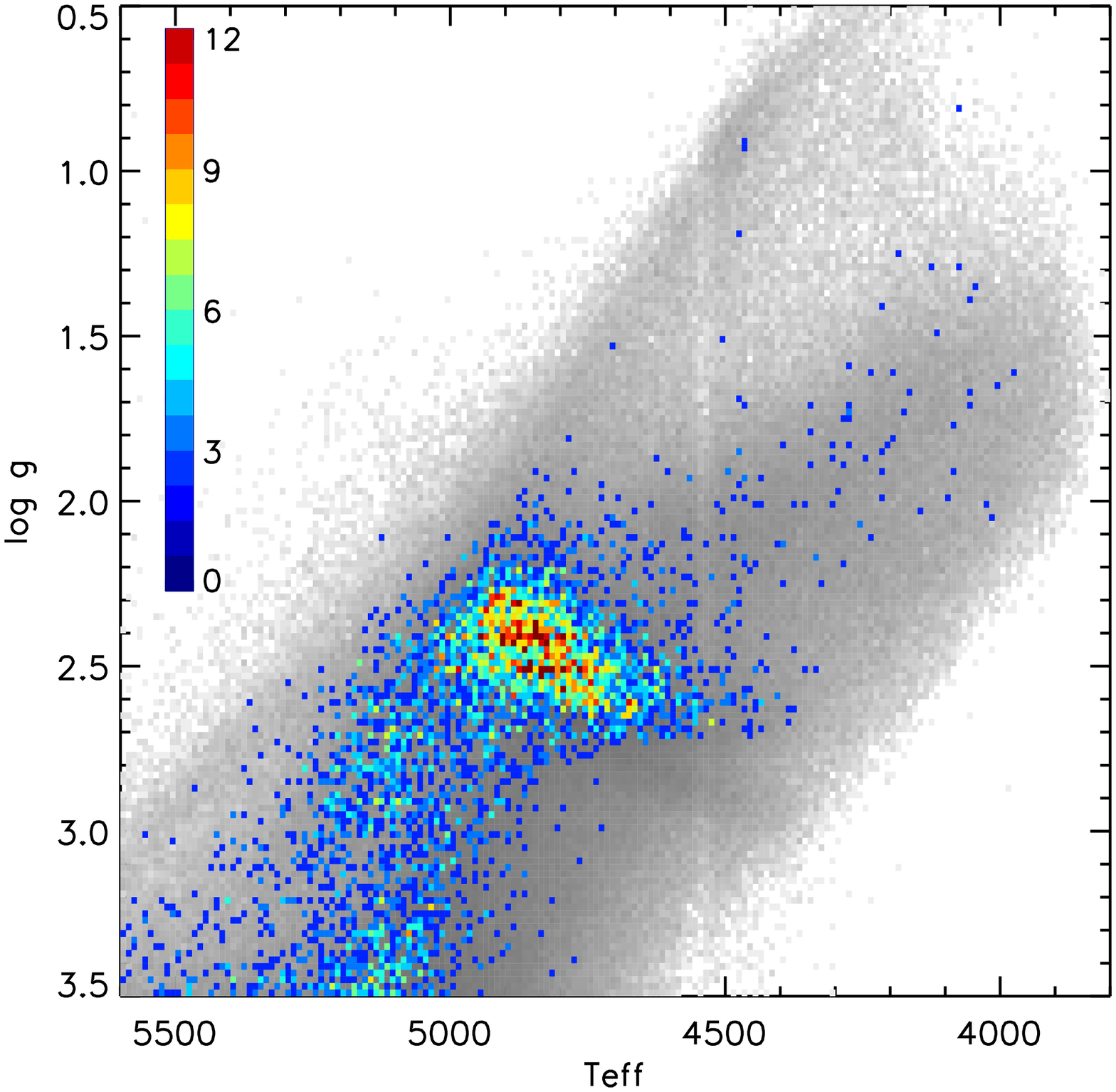}
\centering
\caption{The HR diagram of our Li-rich giant sample (coloured). The all giants are also presented (gray).\label{fig:9}}
\end{figure}

\subsection{Measuring the Li abundances} 

Although the subordinate lines at 6104\,{\AA} and 8126\,{\AA} can be detected for some objects with extremely high Li abundance, they are usually too weak to be detectable in the low-resolution spectra. So the strongest \ion{Li}{1} resonance line at 6708\,\AA\ is used to derive the Li abundances. 

The spectra from LAMOST adopt the vacuum wavelength scale, we converted the vacuum wavelength to air wavelength after corrected the wavelength by the radial velocity. The process to determine the Li abundances follows two steps: 

First, for an object we generated a set of templates with [Li/Fe] various from -3.0 to 6.9, adopted the atmospheric parameters from LASP,  by interpolating the grid of the templates. To check how reliable the interpolated template spectra are, we took three templates from the grids and interpolated their counterparts, and plotted them in Figure \ref{fig:3}. It shows that there is a negligible difference between the original and interpolated ones, which will have no obvious impact on our results.

Second, we calculated the chi-square ($\chi^2$) between each template and the observed spectrum over the wavelength range of 6704\,-\,6712\,{\AA}, which covers the  \ion{Li}{1} resonance line at 6708\,{\AA}. The $\chi^2$ is defined as:
\[{\chi}^2=\sum_{i=1}^{N} \frac{(O_i - T_i)^2}{{\sigma}_i^2}\]
where, ${O_i}$ and ${T_i}$ is the flux of the ${i_{th}}$ point of the observed and the template spectrum, respectively, ${\sigma}_i$ is the error of the observed flux at ${i_{th}}$ pixel, and N is the amount of pixels used in calculation.

Similar to \citet{Xiang2015}, we directly matched the non-normalized observed spectra with the templates, as our targets are giants whose spectra have many absorption lines, it is not easy to estimate the continuum level, this could be worse for the low signal to noise ratio (S/N) spectra. Before calculating the $\chi^2$ value, we corrected the spectral shape between the object and the template on the wavelength range of 6600\,-\,6800\,{\AA} with a third-order polynomial fitting. The $\chi^2$ array was fitted with a Gaussian plus a second-order polynomial to get the minimum $\chi^2$ value, and the corresponding value of [Li/Fe] is determined. Then, A(Li) can be derived.

The \ion{Li}{1} resonance line at 6708\,{\AA} is easily drowned out by noise leading to an invalid result. So we define three values: a) the depth of the \ion{Li}{1} resonance line at 6708\,{\AA} (D); b) the average noise over the wavelength range of 6600-6800\,{\AA} (N); and, c) the standard deviation of the residuals between the object spectrum and the best-matching template (S). 

For each spectrum, we require the following conditions been satisfied:
\begin{center}D $\textgreater$ N \& D $\textgreater$ S\end{center}

The rationale of these two constraints is that the \ion{Li}{1} resonance line should be strong enough in order to affirm the reliability. We automatically eliminate the invalid targets and a small percentage of giants with A(Li) $\geq$ 1.5 have been remained ($\sim3.4\%$).

Then we visually checked them one by one carefully, the main considerations are: whether the \ion{Li}{1} resonance line is obviously unaffected by the noise and the spectrum has credible quality, and whether the Li line of observed spectrum is matchable to the best-fitting template. 
We eliminated the unmatched or bad quality spectra, we also inspected whether there are emission lines of \ion{N}{2} around H$_\alpha$ and \ion{S}{2} around Li resonance in order to get rid of the newly formed objects. Particularly, for the extremely strong Li line at 6708\,{\AA}, we checked the other \ion{Li}{1} lines (6104\,{\AA} and 8126\,{\AA}) and the repeated observations if it had any. 

Figure \ref{fig:4} shows several examples of different A(Li). The light blue region is the wavelength range to calculate $\chi^2$ value, the blue dots represent the observed spectrum, and the red solid line is the best-matching template.

\subsection{Error estimation}\label{sec:Error}
The errors of our A(Li) measurements have two aspects: systematic error due to the intrinsic errors in our method and random errors mainly due to the quality of the observed spectra and/or the uncertainties of the stellar parameters. 

\subsubsection{Systematic error} 

The systematic error of our result is estimated by comparing the Li abundance derived from our method to that from the high-resolution (H.Res.) spectra. In our catalog, 59 Li-rich giants are reported by other high-resolution studies \citep[][in prep.]{Anthony-Twarog2013, Martell2013,Li2018,Zhou2018,Yan2018,Kumar2018b,Singh2019b,Zhou2019,Yan2019}. 
We derived A(Li)$_{\rm LAMOST}$ of these objects using the LAMOST spectra and the stellar atmospheric parameters provided in the literature, and we show the detailed information for 34 published stars in Table \ref{chartable1}. In Figure\,\ref{fig:5}, we present the comparison between A(Li)$_{\rm LAMOST}$ and A(Li)$_{\rm H.Res.}$ for all the 59 stars. It shows a good consistency with an offset of 0.09\,{\it dex} and a dispersion of 0.24\,{\it dex} between our measurements and the results derived by high-resolution spectra in the literature. Thus we consider the systematic error of our result is less than 0.1\,{\it dex}.

\subsubsection{Random errors}

In our results, $2,746$ giants have repeated observations, these could be used to estimate the random errors. We plotted the differences in A(Li) between repeated observations as a function of S/N, {$\mathrm{\emph{T}_{eff}}$}, log \emph{g}, [Fe/H] and A(Li) in Figure \ref{fig:6}. In panel (a), all the $2,746$ objects with repeated observations are included, which shows that the random errors are sensitive to the S/N decreasing from 0.3\,{\it dex} to 0.1\,{\it dex} with increasing S/N. In order to avoid the influence of S/N, only $1,118$ objects with high quality data (S/N $\geq$ 200) were used in the rest panels. Panels (b) and (c) show that the random errors increase from 0.1\,{\it dex} to 0.2\,{\it dex} with increasing {$\mathrm{\emph{T}_{eff}}$} or log \emph{g}, which may be due to the lithium line at 6708\,{\AA} is stronger at low {$\mathrm{\emph{T}_{eff}}$} or log \emph{g}. While the random errors have no obvious relation to the [Fe/H] as shown in panel (d), which may be because the strength of the lithium line has no obvious relationship to [Fe/H]. And panel (e) shows that the scatter of the differences of A(Li) remains same when A(Li) goes from 2.0 to 4.5, and slightly larger in the bin of $1.5 \le A(Li) \textless$ 2.0. It is noted that the lithium line at 6708\,{\AA} is strong enough to be detected when A(Li) higher than 2.0. The typical value of random errors is 0.2\,{\it dex}.

\section{Results}\label{sec:RESULT}

The giants with A(Li) $\geq$ 1.5 are usually defined as Li-rich giants. In our results, 10,535 Li-rich giants are identified. Their information is listed in Table \ref{chartable2}, including the LAMOST ID, positions, the stellar atmospheric parameters provided by LASP, A(Li) and the observed date. Figure \ref{fig:7} shows the histograms of the number of our Li-rich giant sample versus {$\mathrm{\emph{T}_{eff}}$}, log \emph{g} and [Fe/H], respectively. For the distribution of temperature there is a peak around 4800\,K, and might be another peak around 5100\,K. There are two clear peaks around log \emph{g}$\sim$2.5\,{\it dex} and 3.5\,{\it dex}, the first is corresponding to the red giant branch and red clump stars. In addition, the distribution of metallicity shows a clear peak around [Fe/H]$\sim$ -0.15\ and a symmetric profile from $\sim$ $-0.8$ to $+0.5$\,{\it dex}. For the stars with metallicity lower than $-0.8$\,{\it dex}, there seems to be a second peak in the range of $-1.5$ to $-1.0$\,{\it dex}. 
Figure\,\ref{fig:8} shows that the number declines with increasing A(Li). It is noticeable that the number distribution of 1.5\,{\it dex} $\leq$ A(Li) $\leq$ 1.7\,{\it dex} is against the overall trend, this could be because the lithium line is too weak to be detected on the low-resolution spectra when A(Li) is smaller than 1.7\,{\it dex}. Our sample stars in the HR-diagram were displayed with a background of all giant sample in Figure\,\ref{fig:9} , and the two group stars around log \emph{g} of 2.5\,{\it dex} and 3.5\,{\it dex} can also be found.

\section{SUMMARY}\label{sec:SUMMARY}

In this work, we search for Li-rich giants from the LAMOST low-resolution spectra and find $10,535$ Li-rich giants with A(Li) $\geq$ 1.5\,{\it dex}, which is 1.29\% of the all giants in our sample.

We developed a method to derive Li abundance for giants from the low-resolution spectra based on template-matching. We estimate that the systematic error is $\sim 0.1$\,{\it dex} and the random error is around $0.2$\,{\it dex}

The number distribution of our sample in temperature shows two peaks around 4800\,K and 5100\,K, respectively. There are also two clear peaks around 2.5\,{\it dex} and 3.5\,{\it dex} in log \emph{g}. We found a symmetric distribution in the metallicity range of $-0.8$ to $+0.5$\,{\it dex}, while there seems a second peak around $\sim -1.25$\,{\it dex}. As expected, we found that there is a decline of number density with increasing Li abundances.


This is the largest Li-rich giant sample up to date which will help us to investigate the lithium evolution in evolved stars in further work. In the paper \uppercase\expandafter{\romannumeral2}, we will analysis the properties of our Li-rich giant sample from following aspects: the rotation velocity, infrared excess, stellar population and evolutionary stage, etc.\\

\emph{Acknowledgements.} We thank the anonymous referee for his comments which improved this paper. We are grateful to Prof. Chao Liu and Dr. Hao Tian for providing the giant sample. We thank the support from the Key Research Program of the Chinese Academy of Sciences under grant No.XDPB09-02, the National Natural Science Foundation of China under grant Nos. 11833006, 11603037, 11473033, 11973052 and International partnership program's Key foreign cooperation project (No. 114A32KYSB20160049), Bureau of International Cooperation, Chinese Academy of Sciences. H.-L.Y. acknowledges supports from Youth Innovation Promotion Association, CAS and The LAMOST FELLOWSHIP that is supported by Special Funding for Advanced Users, budgeted and administrated by Center for Astronomical Mega-Science, Chinese Academy of Sciences (CAMS). This research is supported by the Astronomical Big Data Joint Research Center, co-founded by the National Astronomical Observatories, Chinese Academy of Sciences and Alibaba Cloud. Guoshoujing Telescope (the Large Sky Area Multi-Object Fiber Spectroscopic Telescope LAMOST) is a National Major Scientific Project built by the Chinese Academy of Sciences. Funding for the project has been provided by the National Development and Reform Commission. LAMOST is operated and managed by the National Astronomical Observatories, Chinese Academy of Sciences.

\begin{deluxetable*}{crrrcrr}
\tablecaption{Information of the Li-rich giants reported by previous works.\label{chartable1}}
\tablewidth{700pt}
\tablehead{
\colhead{ID} & \colhead{{$\mathrm{\emph{T}_{eff}}$}} & 
\colhead{log \emph{g}} & \colhead{[Fe/H]} & 
\colhead{A(Li)$_{\rm LAMOST}$} & \colhead{A(Li)$_{\rm H.Res.}$} & \colhead{Reference}\\
\colhead{} & \colhead{(K)} & \colhead{({\it dex})} & \colhead{({\it dex})} & 
\colhead{({\it dex})} & \colhead{({\it dex})} & \colhead{}
} 
\startdata
  NGC6819-W007017     & 4636  & 2.72  & 0.09  & 2.4  & 2.3  &  \citet{Anthony-Twarog2013} \\
  SDSS J1310-0012     & 4550  & 1.0   & -1.54 & 2.6  & 2.15 &  \citet{Martell2013}  \\ 
  SDSS J0652+4052     & 4900  & 2.9   & 0.04  & 3.4  & 3.3  &  \citet{Martell2013}  \\ 
  SDSS J2353+5728     & 5025  & 3.0   & 0.23  & 3.5  & 3.1  &  \citet{Martell2013}  \\ 
  SDSS J0304+3823     & 5125  & 2.6   & -0.2  & 2.6  & 2.4  &  \citet{Martell2013}  \\ 
 LAMOST J0714+1600          & 5179  & 2.4   & -2.16 & 2.5  &  2.42 &  \citet{Li2018} \\
 LAMOST J0302+1356          & 5206  & 2.3   & -1.74 & 2.5 & 2.34 &  \citet{Li2018} \\
 LAMOST J2146+2732          & 5243  & 2.75  & -1.73 & 3.2 & 2.85 &  \citet{Li2018} \\
  TYC 3251-581-1      & 4670  & 2.3   & -0.09 & 4.1 & 3.68 &  \citet{Zhou2018} \\
  TYC 429-2097-1      & 4696  & 2.25  & -0.36 & 4.6 & 4.63 &  \citet{Yan2018} \\
  KIC2305930          & 4750  & 2.38  & -0.5  & 4.3 & 4.2 &  \citet{Kumar2018b} \\
  KIC12645107         & 4850  & 2.62  & -0.2  & 3.4 & 3.24 &  \citet{Kumar2018b} \\
  TYC 1751-1713-1     & 4830  & 2.58  & -0.25 & 4.2 & 4.15 &  \citet{Singh2019a} \\
  J024710.97+432606.0 &    4315 &    2.18 &    -0.16 &    3.6&     3.24  &  \citet{Zhou2019} \\  
  J055908.81+120339.7 &    4920 &    2.77 &    -0.37 &    4.3&     3.89  &  \citet{Zhou2019} \\ 
  J060649.27+212504.9 &    5188 &    3.16 &    -0.32 &    2.6&     2.53  &  \citet{Zhou2019} \\  
  J064934.47+170424.2 &    5004 &    3.27 &    -0.28 &    4.2&     4.07  &  \citet{Zhou2019} \\  
  J074051.22+241938.3 &    4986 &    2.72 &    -0.17 &    3.9&     4.08  &  \citet{Zhou2019} \\  
  J170124.77+144913.0 &    4796 &    2.75 &    -0.14 &    3.8&     3.51  &  \citet{Zhou2019} \\  
  J011727.43+461528.3 &    4971 &    2.67 &    -0.15 &    3.2&     3.05  &  \citet{Zhou2019} \\  
  J225902.66+054256.2 &    4514 &    2.15 &    -0.1  &    3.1&     3.25  &  \citet{Zhou2019} \\  
  J235043.31+361105.7 &    4716 &    1.71 &    -0.58 &    2.1&     2.31  &  \citet{Zhou2019} \\  
  J071813.82+500452.6 &    4529 &    2.26 &    0.02  &    2.8&     2.62  &  \citet{Zhou2019} \\  
  J072619.82+295808.2 &    4605 &    1.81 &    -0.34 &    3.4&     2.96  &  \citet{Zhou2019} \\  
  J072840.88+070147.4 &    4608 &    1.6  &    -0.28 &    2.6&     2.47  &  \citet{Zhou2019} \\  
  J085929.54+005654.2 &    4018 &    0.62 &    -0.47 &    1.8&     2.18  &  \citet{Zhou2019} \\  
  J103249.02+143714.8 &    5072 &    2.79 &    -0.37 &    3.5&     3.48  &  \citet{Zhou2019} \\  
  J110236.56+133610.3 &    4895 &    2.61 &    -0.35 &    2.3&     2.18  &  \citet{Zhou2019} \\  
  J122234.29+321817.2 &    4430 &    2.18 &    0.08  &    3.8&     4.03  &  \citet{Zhou2019} \\  
  J122525.23+071638.0 &    4764 &    2.16 &    -0.19 &    2.1&     2.06  &  \citet{Zhou2019} \\  
  J132315.71+034347.4 &    4189 &    1.63 &    0.04  &    1.7&     1.85  &  \citet{Zhou2019} \\  
  J143038.38+532629.5 &    4133 &    1.22 &    -0.45 &    1.7&     1.7   &  \citet{Zhou2019} \\  
  J153707.04+182421.0 &    4722 &    2.11 &    -0.06 &    2.6&     2.52  &  \citet{Zhou2019} \\
  J161035.91+331604.8 &    4113 &    1.27 &    -0.79 &    2.8&     2.41  &  \citet{Zhou2019} \\ \enddata    
  \end{deluxetable*}

\begin{deluxetable*}{cccccrcc}
\tablecaption{Information of the Li-rich giants of our sample.\label{chartable2}}
\tablewidth{700pt}
\tablehead{
\colhead{LAMOST ID} & \colhead{R.A.} & \colhead{Decl.} & \colhead{{$\mathrm{\emph{T}_{eff}}$}} & 
\colhead{log \emph{g}} & \colhead{[Fe/H]} & 
\colhead{A(Li)} &  \colhead{DATE}  \\ 
\colhead{} & \colhead{h:m:s (J2000)} & \colhead{d:m:s (J2000)} & \colhead{(K)} & \colhead{({\it dex})} & \colhead{({\it dex})} & 
\colhead{({\it dex})} & \colhead{}
} 
\startdata
LAMOST J000001.30+494500.7 & 00:00:01.30  &   +49:45:00.7 & 4439 &   2.5 &   0.5 & 2.2 &  2014-10-06 \\
LAMOST J000005.50+454110.6 & 00:00:05.50  &   +45:41:10.6 & 4803 &   2.4 &  -0.1 & 3.4 &  2015-10-14 \\
LAMOST J000007.78+410505.4 & 00:00:07.78  &   +41:05:05.4 & 5259 &   3.3 &   0.2 & 2.7 &  2014-12-18 \\
LAMOST J000022.92+544825.2 & 00:00:22.92  &   +54:48:25.2 & 4906 &   2.4 &  -0.4 & 4.1 &  2014-11-20 \\
LAMOST J000036.02+273038.9 & 00:00:36.02  &   +27:30:38.9 & 4958 &   2.4 &  -0.8 & 2.7 &  2016-12-10 \\
LAMOST J000041.35+585002.3 & 00:00:41.35  &   +58:50:02.3 & 4689 &   2.7 &   0.1 & 3.5 &  2014-11-20 \\
LAMOST J000048.98+092600.9 & 00:00:48.98  &   +09:26:00.9 & 4979 &   3.2 &  -0.4 & 1.6 &  2016-12-16 \\
LAMOST J000108.96+072932.9 & 00:01:08.96  &   +07:29:32.9 & 4731 &   2.5 &  -0.2 & 3.6 &  2016-12-16 \\
LAMOST J000119.92+082335.9 & 00:01:19.92  &   +08:23:35.9 & 4801 &   2.3 &  -0.5 & 2.4 &  2016-12-16 \\
LAMOST J000133.56+554937.3 & 00:01:33.56  &   +55:49:37.3 & 4904 &   2.4 &  -0.0 & 1.6 &  2014-11-20 \\
LAMOST J000143.05+254549.5 & 00:01:43.05  &   +25:45:49.5 & 4655 &   2.6 &   0.2 & 1.6 &  2016-12-10 \\
LAMOST J000151.65+265848.4 & 00:01:51.65  &   +26:58:48.4 & 5071 &   2.5 &  -0.5 & 4.7 &  2016-12-10 \\
LAMOST J000156.01+372623.2 & 00:01:56.01  &   +37:26:23.2 & 4519 &   2.2 &  -0.4 & 3.7 &  2012-11-30 \\
LAMOST J000201.61+445049.1 & 00:02:01.61  &   +44:50:49.1 & 4948 &   2.5 &  -0.4 & 3.6 &  2015-10-14 \\
LAMOST J000205.10+384906.2 & 00:02:05.10  &   +38:49:06.2 & 4937 &   3.1 &  -0.1 & 1.6 &  2014-10-05 \\
LAMOST J000206.98+472520.2 & 00:02:06.98  &   +47:25:20.2 & 4640 &   2.9 &   0.3 & 1.9 &  2013-10-30 \\
LAMOST J000211.20+532701.4 & 00:02:11.20  &   +53:27:01.4 & 4830 &   2.4 &  -0.3 & 3.0 &  2014-10-06 \\
LAMOST J000227.22+493429.9 & 00:02:27.22  &   +49:34:29.9 & 3989 &   1.4 &  -0.2 & 2.7 &  2017-10-16 \\
LAMOST J000230.61+582629.3 & 00:02:30.61  &   +58:26:29.3 & 5139 &   2.8 &   0.1 & 2.5 &  2014-11-20 \\
LAMOST J000242.92+435331.3 & 00:02:42.92  &   +43:53:31.3 & 5194 &   2.5 &  -0.4 & 1.6 &  2014-12-18 \\
... & ... & ... & ... &  ... & ... & ... &  ...\\\enddata
\end{deluxetable*}

\end{document}